\documentclass[%
 aip,
 cha,
 amsmath,amssymb,
 preprint,%
]{revtex4-1}

\usepackage{graphicx}
\usepackage{dcolumn}
\usepackage{bm}

\usepackage[utf8]{inputenc}
\usepackage[T1]{fontenc}
\usepackage{mathptmx}
\usepackage{wasysym}

\begin{document}

\preprint{AIP/CHA19-FT-01188}

\title[Chaotic transport of navigation satellites]{Chaotic transport of navigation satellites}

\author{Ioannis Gkolias}
 \email{ioannis.gkolias@polimi.it}
 \affiliation{Department of Aerospace Science and Technology, Politecnico di Milano, Milan, 20156, Italy.}

\author{J\'er\^ome Daquin}
\affiliation{Department of Mathematics, University of Padova, Padova, 35131, Italy.}

\author{Despoina K.~Skoulidou}
\affiliation{Department of Physics, Aristotle University of Thessaloniki, Thessaloniki, 54124, Greece.} 

\author{Kleomenis Tsiganis}
\affiliation{Department of Physics, Aristotle University of Thessaloniki, Thessaloniki, 54124, Greece.}

\author{Christos Efthymiopoulos}
\affiliation{Research Center for Astronomy and Applied Mathematics, Academy of Athens, Athens, 11527, Greece.}
\affiliation{Department of Mathematics, University of Padova, Padova, 35131, Italy.}

\date{\today}

\begin{abstract}
Navigation satellites are known from numerical studies to reside in a dynamically sensitive environment, which may be of profound importance for their long-term sustainability. We derive the fundamental Hamiltonian of GNSS dynamics and show analytically that near-circular trajectories lie in the neighborhood of a Normally Hyperbolic Invariant Manifold (NHIM), which is the primary source of hyperbolicity. Quasi-circular orbits escape through chaotic transport, regulated by the NHIM's stable and unstable manifolds, following a power-law escape time distribution $P(t) \sim t^{-\alpha}$, with $\alpha \sim 0.8 - 1.5$. Our study is highly relevant for the design of satellite disposal trajectories, using manifold dynamics.
\end{abstract}

\maketitle

\begin{quotation}
Global Navigation Satellite Systems (GNSS) reside on quasi-circular Medium Earth Orbits (MEO), largely inclined with respect to the Earth's equator. Resonant gravitational interactions with the Moon and the Sun can significantly increase GNSS eccentricities on decadal time-scales, leading to `Earth-crossing' orbits, but this depends sensitively on initial conditions, as shown in numerical studies. Here we derive the fundamental Hamiltonian of GNSS dynamics and show analytically that operational trajectories lie in the neighborhood of a normally hyperbolic invariant manifold. Chaos becomes prominent precisely at {\it Galileo} altitudes, where two lunisolar resonances cross; this is a consequence of the exact value of the well-known period of precession of the Moon's orbit about the ecliptic  (18.6~years). Inside the tangle of stable and unstable manifolds that encompasses circular orbits, short-lived trajectories alternate with long-lived ones in a fractal pattern and transport is characterized by a power-law distribution of escape times. As shown here, knowledge of the local manifold dynamics can be used to target the `fast-escaping' trajectories. Thus, apart from explaining a long-known phenomenology, our study opens a new path for the efficient design of end-of-life (EoL) disposal strategies, which is important for GNSS sustainability.
\end{quotation}

\section{Introduction}
GNSS are constellations of $\sim 30$ satellites each, residing on almost circular (eccentricities are $e\sim 10^{-4}$), inclined MEO. They include the Russian GLObal NAvigation Satellite System (GLONASS)  (semi-major axis $a=25,500~$km, inclination $i=65^{\circ}$), GPS ($a=26,560~$km, $i=55^{\circ}$), Beidou ($a=27,900~$km, $i=55^{\circ}$) and Galileo ($a=29,600~$km, $i=56^{\circ}$) systems. Constellation design requires multi-objective optimization, Earth coverage and cost being the primary constraints. For MEO altitudes, optimal solutions yield $i\sim 52-58~$degrees with respect to the Earth's equator \cite{Abbondanza2001,mozo2001}; for GLONASS, sufficient coverage at high latitudes requires $i\sim 62-68~$degrees. \\

Long-term sustainability of GNSS calls for the development of efficient EoL disposal strategies that will safeguard the constellations from defunct `debris' \cite{Liou2006,Rossi2008,Alessi2014,Rosengren2017,Armellin2018,Skoulidou2019}. However, the chosen optimal inclinations induce complications, as they coincide with the phase-space loci of gravitational lunisolar resonances \cite{Cook1962,Hughes1980,Hughes1981,Breiter2001,Ely1997}. The celebrated {\it Lidov-Kozai} resonances \cite{Lidov1962,Kozai1962,Breiter2001}, occurring for all values of $a$ but at specific and nearly fixed values of $i$, are commensurabilities between the precession rates of the argument of the perigee, $g=\omega$, and the right ascension of the ascending node, $h=\Omega$, of a satellite's orbit. The relevant terms of the perturbing potential can be identified using Legendre-type expansions and analytically tractable, averaged (over short-periods) Hamiltonians can be defined \cite{Giacaglia1974,Lane1989,Lara2014}. At $i=56^{\circ}$, the dominant term is associated with the $2 \dot{g} + \dot{h}=0 $ resonance \cite{Stefanelli2015,Celleti2016} ($\mathcal{R}_{2g+h}$), which is the focus of this study. \\

Several numerical studies have highlighted the significant eccentricity boost received by MEOs in this resonance  \cite{Chao2004,Rossi2008,Deleflie2011,Alessi2016,Skoulidou2019} and the emergence of chaotic transport, associated with the precession of the lunar nodes \cite{Rosengren2015,Daquin2016,Gkolias2016} and with the influence of multiple resonances \cite{Rosengren2017,Breiter2001b}. Eccentricity growth offers a natural disposal solution, as lowering of the satellite's perigee can lead to atmospheric re-entry. Previous studies suggest that this mechanism is very sensitive to the choice of initial conditions \cite{Rosengren2015,Daquin2016,Gkolias2016}. Since chaos prevents us from accurately predicting when a defunct satellite will actually evacuate the $e \approx 0$ operational zone, understanding the mechanism of chaotic transport and identifying possible ways of controlling it, is important to EoL strategies design. \\

\section{Analytical theory}

MEO satellite dynamics can be modelled by the following Hamiltonian

\begin{equation}
\label{eq:fullmodel}
\mathcal{H} = \mathcal{H}_{Kep} + \mathcal{H}_{J_2} + \mathcal{H}_{LS},
\end{equation}
where
\begin{eqnarray}
\mathcal{H}_{kep} &=& \frac{v^2}{2} - \frac{\mu_{\oplus}}{r},  \\
\mathcal{H}_{J_2} &=& \frac{R_\oplus^2 J_2 \mu_{\oplus}  \left(3 \sin^2\phi-1\right)}{2 r^3}, \\ 
\mathcal{H}_{LS} &=& - \frac{\mu_{\leftmoon}}{r_{\leftmoon}} \left(\frac{r_{\leftmoon}}{||\bf{r}-\bf{r}_{\leftmoon}||} - \frac{\mathbf{r}\cdot \bf{r}_{\leftmoon}}{r_{\leftmoon}^2} \right) - \frac{\mu_{\odot}}{r_{\odot}} \left(\frac{r_{\odot}}{||\bf{r}-\bf{r}_{\odot}||} - \frac{\mathbf{r}\cdot \bf{r}_{\odot}}{r_{\odot}^2} \right).
\end{eqnarray}
\noindent
$\mathcal{H}_{Kep}$ corresponds to the Kepler problem, with $\mu_{\oplus}$ the gravitational parameter of the Earth, and $r$, $v$ being the geocentric distance and velocity of the satellite. $\mathcal{H}_{J_2}$ is the perturbation caused by the Earth's oblateness, with $J_2$ the oblateness parameter, $R_{\oplus}$ the mean equatorial radius of the Earth, and $\phi$ the geocentric latitude of the satellite. $\mathcal{H}_{LS}$ is the lunisolar perturbation, with $\bf{r}_{\leftmoon}, \bf{r}_{\odot}$ the geocentric vectors of the Moon and the Sun respectively, $r_{\leftmoon}$,$r_{\odot}$ the corresponding geocentric distances and $\mu_{\leftmoon}$,$\mu_{\odot}$ Moon's and Sun's gravitational parameters. 

In celestial mechanics, following the Keplerian notation, we express the Hamiltonian in terms of canonical functions of the orbital elements. A Legendre-type expansion of $\mathcal{H}_{LS}$ up to quadrupolar terms in the geocentric distances of the Moon and the Sun is performed and $\mathcal{H}$ is averaged over the mean motions of all objects. Thus, the {\it secular Hamiltonian} reduces to a time-dependent, two degrees-of-freedom model. The Delaunay momentum $L=\sqrt{\mu_{\oplus} a}$ is preserved, while time enters through the precession of the ecliptic lunar node, $\Omega_{\leftmoon} \approx \Omega_{\leftmoon,0} + n_{\Omega_{\leftmoon}} \, t$ \cite{Cook1962}, with frequency $n_{\Omega_{\leftmoon}}$ that corresponds to the known lunar nodal precession cycle of $2\pi /| n_{\Omega_{\leftmoon}}| \simeq 18.6~$years. We adopt the value $i_{\leftmoon}=5^{\circ}.15$ for the Moon's inclination to the ecliptic plane.\\

We apply the canonical transformation defined in [\onlinecite{Breiter2001}] to resonant variables $(J_R,J_F,u_R,u_F)$, appropriate for the resonant argument $u_R \,= \, -g \, -h/2$, through $(G,H,g,h)  \, = \, ( L - J_R, \, L - J_R - I_F, \, u_F /2 - u_R, - u_F )$, where $(G,H)=(\sqrt{\mu_{\oplus} a(1-e^2)},\, \sqrt{\mu_{\oplus} a(1-e^2)} \cos i)$ are the expressions of the norm and the $z$-component of the satellite's angular momentum in orbital elements. An additional Taylor expansion around the unperturbed, exact resonance $(J_R, I_F) \,= \,(0,I_F^{\star})$, with $I_F^{\star} = \sqrt{\mu_{\oplus} a} (\cos{i_{\star}} - 1 )$ and $i_{\star} = 56^{\circ}.06$, followed by a transformation to non-singular Poincar\'e variables $(X,Y) = (\sqrt{2 J_R} \sin{u_R} , \sqrt{2 J_R} \cos{u_R} )$ leads to the final reduced Hamiltonian

\begin{equation}
\label{eq:hsecular}
\bar{\mathcal{H}} = \mathcal{H}_{R} + \mathcal{H}_{CM} + \mathcal{H}_{C},
\end{equation}

\noindent
where,  
\begin{equation}
\label{eq:HR}
\mathcal{H}_{R} =  c_{20} X^2 +  c_{02} Y^2 +  c_{22} X^2 Y^2 + c_{40} X^4 + c_{04} Y^4 + \ldots,
\end{equation}
$\mathcal{H}_{CM} = \mathcal{H}_{CM,0} + \mathcal{H}_{CM,1}$ with
\begin{eqnarray}
\label{eq:HCM}
\mathcal{H}_{CM,0} &=&  b_{10} J_F +  b_{20} J_F^2 +  b_{01} \cos{u_F} + b_{02} \cos{2 u_F} + \ldots, \\ 
\mathcal{H}_{CM,1} &=&  n_{\Omega_{\leftmoon}} J_{\leftmoon} +  d_{21} \cos{(2 u_F + \Omega_{\leftmoon})} + d_{11} \cos{(u_F + \Omega_{\leftmoon})} + \ldots,
\end{eqnarray}
and
\begin{equation}
\label{eq:HC}
\mathcal{H}_{C} =  c_{120} J_F X^2 +  c_{102} J_F Y^2 + \ldots .
\end{equation}

\noindent
where $J_F = I_F-I_F^{\star}$, $X\sim - e \sin(g+h/2)$ and $Y \sim e \cos(g+h/2)$ are $\mathcal{O}(e)$ and $J_{\leftmoon}$ is a dummy action conjugate to $\Omega_{\leftmoon}$. The coefficients in Eqs.~(\ref{eq:HR})-(\ref{eq:HC}) are expressed in terms of the relevant physical and dynamical parameters in Table~\ref{tab:ceoffs}.\\

\begin{table}[thb]
\caption{\label{tab:ceoffs}%
Coefficients of the leading terms in $\bar{\mathcal{H}}$. The values of the parameters are $\mu_{\odot}=1.32712\cdot10^{11}~km^3/s^2,\mu_{\leftmoon}=4902.8~km^3/s^2,\mu_{\oplus}=398600~km^3/s^2$, $J_2=1.082 \cdot 10^{-3}$, $R_{\oplus}=6378.1~$km, $r_{\odot}=1.49579\cdot 10^8~$km, $r_{\leftmoon}=384157~$km. $c_{i_{\leftmoon}},s_{i_{\leftmoon}}$ are the sine and cosine of the inclination of the Moon to the ecliptic $i_{\leftmoon}=5^{\circ}.15$, $c_{\epsilon},s_{\epsilon}$ the sine and cosine of the obliquity of the ecliptic $\epsilon=23^{\circ}.44$, $s_{i_{\star}},c_{i_{\star}}$ the sine and cosine of $i_{\star} = 56^{\circ}.06$ and $n=\sqrt{\mu_{\oplus}/a^3}$ the mean motion of the satellite.    
}
\begin{ruledtabular}
\begin{tabular}{ccc}
\textrm{Terms}&
\textrm{Coefficients}&
\textrm{Values}\\
\colrule
$X^2, Y^2$ & $c_{20},c_{02}$ & $\mp \frac{ 15 \mu_{\odot} (1 + c_{i_{\star}}) c_{\epsilon} s_{i_{\star}} s_\epsilon}{16 n r_{\odot}^3} \pm \frac{15 \mu_{\leftmoon} (1 + c_{i_{\star}}) c_{\epsilon} s_{i_{\star}} s_\epsilon (3 s_{i_{\leftmoon}}^2-2)}{32 n r_{\leftmoon}^3}$\\
$J_F X^2$, $J_F Y^2$ & $c_{120},c_{102}$ & $\begin{array}{c}  \frac{3 J_2 R_{\oplus}^2 ( 10 c_{i_\star} -1)}{8 a^4} + \frac{\mu_{\odot} ( 18 -54 c_{i_\star} \pm 30 (2 c_{i_\star} -1) c_\epsilon s_{i_\star} s_\epsilon + 27 (3 c_{i_\star}-1) s_\epsilon^2}{32 a^2 n^2 r_{\odot}^3 (c_{i_\star}-1)} \\ + \frac{3 \mu_{\leftmoon} (3 s_{i_{\leftmoon}}^2 -2)(18 c_{i_\star}-6 \pm 10 ( 1 - 2 c_{i_\star}) c_\epsilon s_{i_\star} s_\epsilon + 9 (1 - 3 c_{i_\star}) s_\epsilon^2)}{64 a^2 n^2 r_{\leftmoon}^3 (c_{i_\star}-1)} \end{array}$ \\
$J_F$ & $ b_{10}$ & $ \frac{3 J_2 n R_{\oplus}^2 c_{i_\star}}{2 a^2} + \frac{ 3 \mu_{\odot} c_{i_\star} (2 - 3 s_\epsilon^2)}{
 8 n r_{\odot}^3} + \frac{3 \mu_{\leftmoon} c_{i_\star} ( 3  s_{i_{\leftmoon}}^2 - 2) ( 3 s_\epsilon^2 - 2)}{16 n r_{\leftmoon}^3}$ \\
$J_F^2$ & $ b_{20}$ & $-\frac{3 J_2 R_{\oplus}^2}{4 a^4}+ \frac{ 3 \mu_{\odot} ( 3 s_{i_{\leftmoon}}^2 - 2 ) }{16 a^2 n^2 r_{\odot}^3} - \frac{ 3 \mu_{\leftmoon} (3 s_{i_{\leftmoon}}^2 - 2) ( 3 s_\epsilon^2 - 2)}{32 a^2 n^2 r_{\leftmoon}^3}$ \\
$\cos{u_F}$ &  $b_{01}$ & $ - \frac{3 a^2 \mu_{\odot} c_{i_\star} c_\epsilon s_{i_\star} s_\epsilon}{4 r_{\odot}^3} + \frac{ 3 a^2 \mu_{\leftmoon} c_{i_\star} s_{i_\star} c_\epsilon s_\epsilon (3 s_{i_{\leftmoon}}^2 - 2) }{8r_{\leftmoon}^3}$ \\
$\cos{2u_F}$ & $ b_{02}$ &  $\frac{3 a^2 \mu_{\odot} (c_{i_\star}^2-1)s_\epsilon^2}{16 r_{\odot}^3} - \frac{3 a^2 \mu_{\leftmoon} (c_{i_\star}^2-1)(3 s_{i_{\leftmoon}}^2-2) s_\epsilon^2}{32 r_{\leftmoon}^3}$ \\
$\cos(2 u_F + \Omega_{\leftmoon})$ & $ d_{21}$ & $ \frac{3 a^2 \mu_{\leftmoon} (c_{i_\star}^2 -1) c_{i_{\leftmoon}} (1 + c_\epsilon) s_{i_{\leftmoon}} s_\epsilon}{16 r_{\leftmoon}^3} $\\
$\cos(u_F + \Omega_{\leftmoon})$ & $ d_{11}$ & $ -\frac{3 a^2 \mu_{\leftmoon} c_{i_\star} s_{i_\star} c_{i_{\leftmoon}} s_{i_{\leftmoon}} (1 + c_\epsilon -2 s_\epsilon^2)}{8 r_{\leftmoon}^3} $\\
\end{tabular}
\end{ruledtabular}
\end{table}

Note that $\mathcal{H}_R$ and $\mathcal{H}_C$ are both $\mathcal{O}(e^2)$, while $\mathcal{H}_{CM}$ does not depend on $X,Y$. As a consequence, circular orbits satisfy for all time the invariance equations  $\dot{X}=\dot{Y}= 0 = X = Y$. For these orbits, the evolution of $J_F$ is given by $\mathcal{H}_{CM}=\mathcal{H}_{CM,0}+\mathcal{H}_{CM,1}$, which defines an invariant subset of the phase space of $\bar{\mathcal{H}}$, the \emph{center manifold} (CM). The term `center manifold' here denotes an invariant manifold embedded in the phase space, whose tangent dynamics is neutral. The CM is not associated with an exact equilibrium point of the flow and it is not isoenergetic. Neglecting $\mathcal{H}_{CM,1}$, the CM would be foliated in rotational tori, describing small oscillations with amplitude equal to the inclination of the Laplace plane \cite{Kudielka1997,Tremaine2009}, $\Delta i\approx 0^{\circ}.5-1^{\circ}.7$ for MEO satellites. However, for $a\approx 29,930~$km, which is slightly above the {\it Galileo} altitude, $\cos(2 u_F + \Omega_{\leftmoon})$ becomes near-resonant ($\mathcal{R}_{2h-\Omega_{\leftmoon}})$ and this can increase significantly $\Delta i$. To lowest order, the Hamiltonian of Galileo dynamics becomes

\begin{equation}
\mathcal{H}_{Gal} = \mathcal{H}_{CM,0} + n_{\Omega_{\leftmoon}} J_{\leftmoon} +  d_{21} \cos{(2 u_F + \Omega_{\leftmoon})}.
\end{equation}

\noindent Defining the slow angle $u_1=u_F+\Omega_{\leftmoon}/2$, we can eliminate $u_F$ using a canonical transformation, such that the Hamiltonian reduces to a pendulum form

\begin{equation}
\label{eq:centerpend}
\bar{\mathcal{H}}_{Gal} = \mathcal{C}_{10} J_1 + \mathcal{C}_{20} J_1^2 + \mathcal{C}_{01} \cos{2 u_1},
\end{equation} 

\noindent 
where $J_1 = J_F + \frac{b_{01}}{b_{10}} \cos{u_F} +  \frac{b_{02}}{b_{10}} \cos{2 u_F} + \mathcal{O} ( J_F ^2)$, $\mathcal{C}_{10} = b_{10} + n_{\Omega_{\leftmoon}}/2$, $\mathcal{C}_{20} = b_{20}$ and $\mathcal{C}_{01} = d_{21}$. The secular variations of $J_F$ can now be approximated by the pendulum solutions. Its average value over a phase torus, $\langle J_F \rangle$, defines an approximate integral of motion, namely a {\it proper} inclination for circular orbits on the CM, given by
\begin{equation}
i_{P} = \arccos{\left( \frac{L - (\langle J_F \rangle + I_F^{\star})}{L} \right)},
\end{equation}

\noindent
where, for librations of $u_1$ 
\begin{equation}
\langle J_F \rangle_{lib} = - \frac{\mathcal{C}_{10}}{2 \mathcal{C}_{20}}, 
\end{equation} 

\noindent
and for circulations  
\begin{equation}
\langle J_F \rangle_{circ} =  \langle J_F \rangle_{lib} \pm \frac{\sqrt{\mathcal{C}_{10}^2+ 4 \mathcal{C}_{20} \, \bar{\mathcal{H}}_{Gal}^0 }} {2 \mathcal{C}_{20} },
\end{equation}
with $\bar{\mathcal{H}}_{Gal}^{0}$ equal to the value of Eq.~(\ref{eq:centerpend}) for a given a set of initial conditions $(X=Y=0, J_{F,0},u_{F,0},u_{m,0})$. Hence, $\langle J_F \rangle$ is a function of the initial conditions $(J_F,u_F,u_m)$ on the CM. This allows us to compute the inclinations range, for which the CM becomes a {\it normally hyperbolic invariant manifold} (NHIM) \cite{Wiggins1994}. Substituting $\langle J_F \rangle$ in the Hamiltonian $\bar{\mathcal{H}}_P = \mathcal{H}_{R} + \mathcal{H}_{C}$, we characterize stability in the neighborhood of the CM ($X=Y=0$), by an approximation based on the eigenvalues of the linearized variations matrix $\mathcal{D}_0$, associated with the flow of $\bar{\mathcal{H}}_P$ 
   
\begin{equation}
\label{eq:stabmat0}
\mathcal{D}_{0} = \left( \begin{array}{cc} 0 &  2 c_{02} +  2 c_{102} \langle J_F \rangle\\  - 2 c_{20} -  2 c_{120} \langle J_F \rangle& 0\end{array} \right).
\end{equation}
Note that $\mathcal{D}_0$ depends on the initial conditions on the CM via $\langle J_F \rangle$. The NHIM corresponds to the subset of points $(J_F,u_F,u_m)$ on the CM, for which the $\langle J_F \rangle$ leads to real eigenvalues of $\mathcal{D}_0$.\\

Chaotic transport is, now, expected to be regulated by the stable and unstable manifolds of the NHIM \cite{Wiggins2001,Naik2019,Dellnitz2005,Jaffe2002}. The motion transversely to the CM is approximately described by $\bar{\mathcal{H}}_P$, with $J_F$ in $\mathcal{H}_{C}$ substituted by $J_F(t) \approx J_1(t) - \frac{b_{01}}{b_{10}} \cos( b_{10} t) - \frac{b_{02}}{b_{10}} \cos(2 b_{10} t)$ and $J_1(t)$ taken from the integrable (\ref{eq:centerpend}). Then, the $J_F X^2$ and $J_F Y^2$ terms in $\mathcal{H}_C$ give $\mathcal{O}(\sin i_{\leftmoon})$ oscillations of the separatrix of $\bar{\mathcal{H}}_P$. In fact, if the Moon `is set' on the ecliptic in a numerical simulation (i.e. $\sin i_{\leftmoon} = 0$), chaotic transport disappears.     

\section{Manifold dynamics and chaotic transport}

\begin{figure}[t]
	\includegraphics[width= \columnwidth]{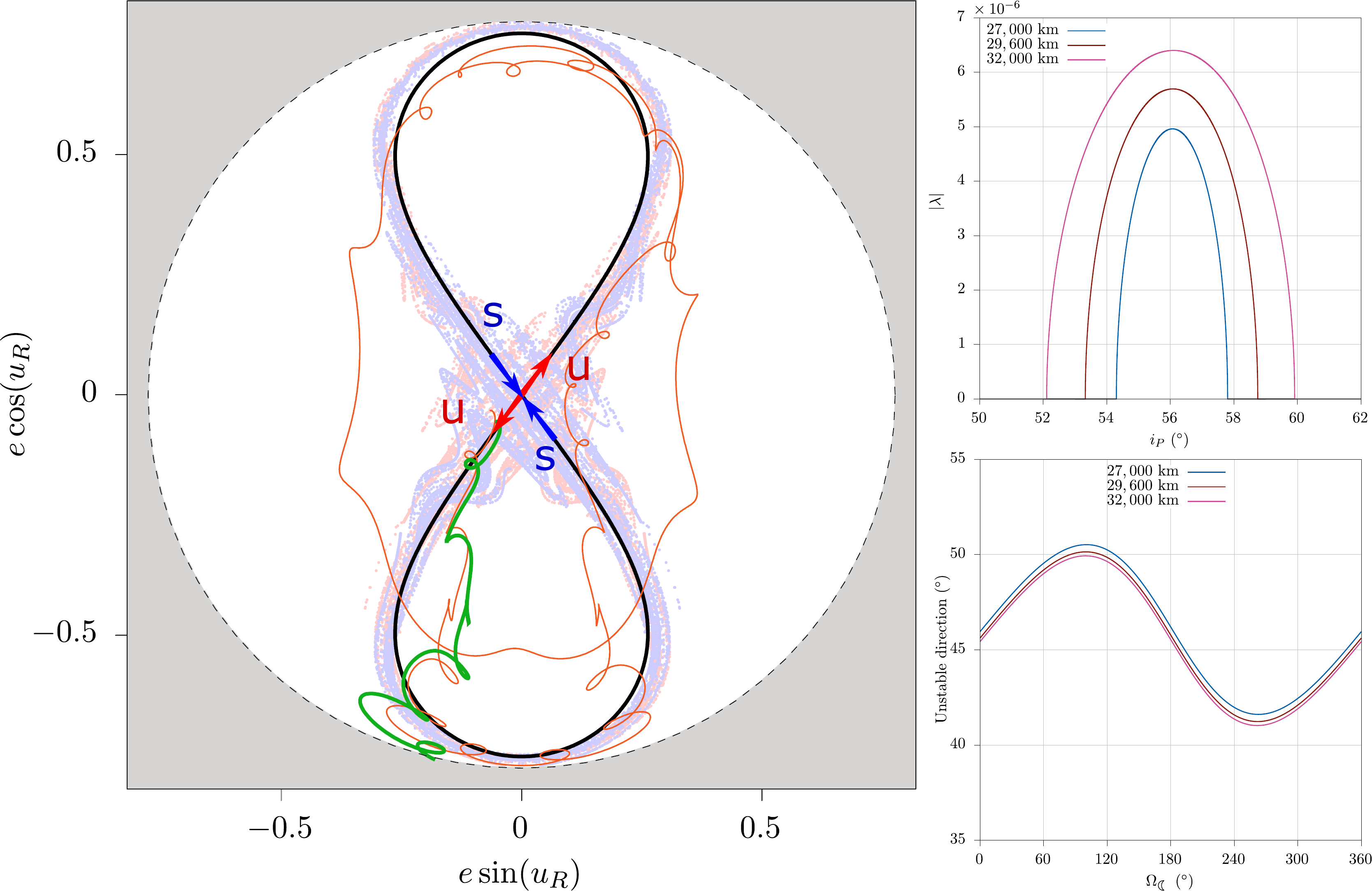}
	\caption{\label{fig:analytical} (left) Composite phase portrait for Galileo dynamics. The separatrix is taken from $\bar{\mathcal{H}}_P$ (black line); stable (S) and unstable (U) eigendirections are shown with blue and red arrows. The light blue/red dotted lines are a numerical realization of parts of the stable and unstable manifolds of the NHIM, computed from a periodic orbit of $\bar{\mathcal{H}}$. Two nearby trajectories of $\mathcal {H}$ are shown, one driven to atmospheric re-entry, following the unstable manifold (green), whereas the second is trapped in a recurrent motion about the center (orange). (top-right) The eigenvalues of $\mathcal{D}_0$, as functions of $i_{P}$, for different altitudes. (bottom-right) Dependence of the unstable eigendirection on the initial phase of the lunar node.}
\end{figure}

As can be seen in Fig.~\ref{fig:analytical}, the eigenvalues of Eq.~(\ref{eq:stabmat0}) are real and the CM is normally hyperbolic for $53^{\circ} \leq i_P \leq 59^{\circ}$, at Galileo altitude. The largest eigenvalue maximizes for $i_{P}=56^{\circ}$, giving an e-folding time of $\approx 36~$years. Close to this maximum, the extent of the separatrix of $\bar{\mathcal{H}}_P$ is $\mathcal{O}(1)$. This is a direct consequence of the near preservation of $J_F$, which results into coupled oscillations of $e$ and $i$, as $G - H - I_{F}^{\star} = J_F \approx 0$. Hence, inclination variations of size $\Delta i$ lead to 

\begin{equation}
\Delta e = \left| \frac{2 \sin{i_{\star}} \Delta i}{\frac{1}{2} - \cos{i_{\star}}} \right|^{1/2} \approx e_{\max},
\end{equation}

\noindent
and quasi-circular orbits can reach $e_{\max} >  0.7$ and become `Earth-crossing' (shaded area in Fig.~\ref{fig:analytical}), as seen in numerical simulations. Similar behavior has also been reported about the $\mathcal{R}_{2g}$ Lidov-Kozai resonance \cite{Breit2007,Gkolias2019}. As discussed below, the trajectories of Eq.~(\ref{eq:fullmodel}) follow closely the manifolds of our double-resonance Hamiltonian (Eq.~\ref{eq:hsecular}), shown in Fig~\ref{fig:analytical}. Note the near-perpendicular intersections of the manifolds close to the origin, which results in a disc of size $\delta e \sim 0.2$ being immersed in the chaotic tangle.\\

\begin{figure}
	\includegraphics[width= \columnwidth]{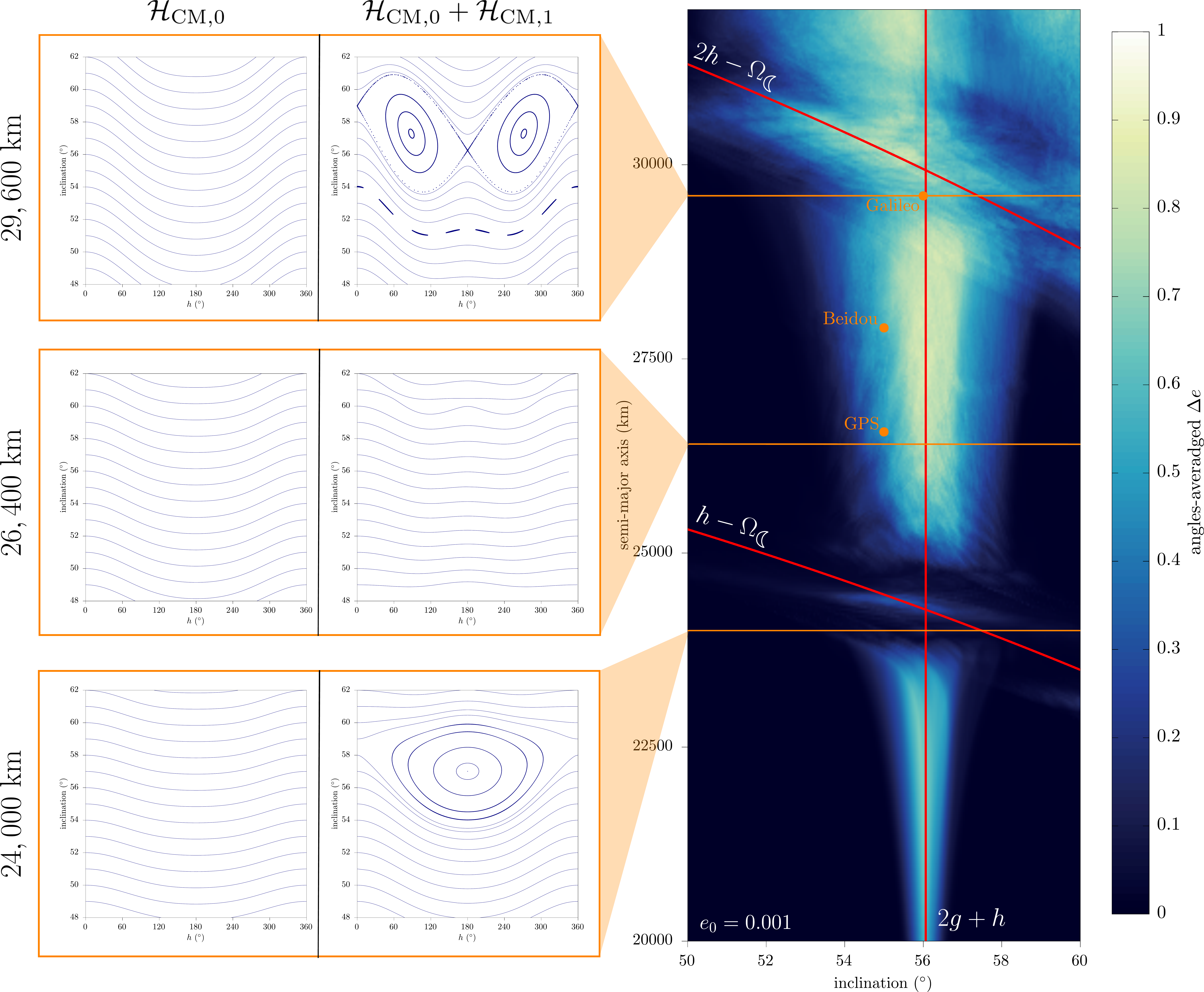}
	\caption{\label{fig:map2} (left) Phase diagrams of $\mathcal{H}_{CM,0}$ (left column) and $\mathcal{H}_{CM,0}+\mathcal{H}_{CM,1}$ (right column) at three altitudes. The difference in these plots clearly marks the importance of the $h-\Omega_{\leftmoon}$ ($24,270~$km) and $2h-\Omega_{\leftmoon}$ ($29,930~$km) lunisolar resonances. As seen also in the corresponding $\Delta e$ map of Eq.~(\ref{eq:fullmodel}) (right), these two resonances actually have opposite effect on the stability of the CM. The positions of the GNSS constellations are also shown (orange circles).}
\end{figure}

Fig.~\ref{fig:map2} is a map of angles-averaged $\Delta e$ \cite{Gkolias2016}, as computed for a dense grid of initial conditions in $(a,i)$, under Eq.~(\ref{eq:fullmodel}). The width, $\Delta i_{res}$, of the high-$\Delta e$ region found at any given altitude, corresponds to the region of real eigenvalues of Eq.~(\ref{eq:stabmat0}). The seemingly uniform increase of $\Delta i_{res}$ with $a$ is interrupted at $29,930~$km, where the $\mathcal{R}_{2h-\Omega_{\leftmoon}}$ resonance crosses the domain of the $\mathcal{R}_{2g+h}$. Its importance is clearly seen in the phase-diagrams of the CM dynamics, attached to the eccentricity variations map. The addition of $\mathcal{H}_{CM,1}$ to $\mathcal{H}_{CM}$ leads to the appearance of a separatrix, which results to substantial increase of $\Delta i$, as opposed to $\mathcal{H}_{CM,0}$; in Fig.~\ref{fig:analytical}, this corresponds to a large area of the $(X,Y)$ plane occupied by the stable and unstable manifolds of the NHIM. Note that, similarly to the Galileo $\mathcal{R}_{2h-\Omega_{\leftmoon}}$ resonance at $a=29,930~$km, the $\mathcal{R}_{h-\Omega_{\leftmoon}}$ resonance becomes important at $24,270~$km. However, a similar analysis as above shows that the $\cos{(u_F+\Omega_{\leftmoon})}$ harmonic actually restores elliptic stability of the CM in two zones, immediately above and below $24,270~$km; this is confirmed by computing the eigenvalues of the $\mathcal{R}_{h-\Omega_{\leftmoon}}$-dependent Hamiltonian analogous to $\bar{\mathcal{H}}_{P}$. \\

\begin{figure}
	\includegraphics[width= \columnwidth]{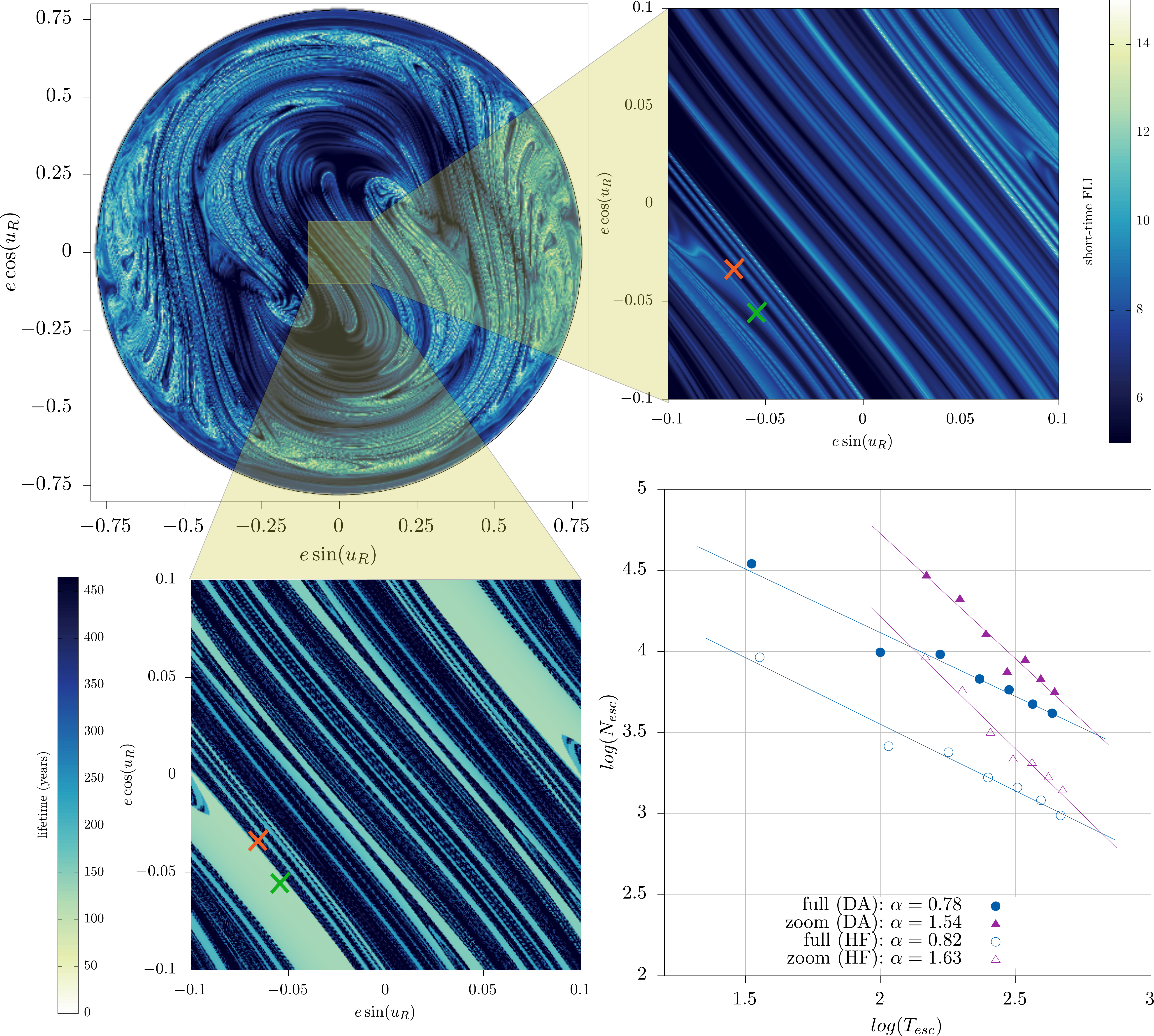}
	\caption{\label{fig:numerical} (top-left) Short-time FLI map of the entire grid, depicting the stable manifold of the NHIM. Attached, a zoom of the central region (top-right), whose $T_{esc}$ map is also shown (bottom-left). Green and orange crosses mark the initial conditions of the two trajectories of the same color in Fig.~\ref{fig:analytical}. The distributions of $T_{esc}$ follow power-laws (bottom-right).}
\end{figure}

The effect of the manifolds structure on the escape dynamics of Galileo satellites is shown in Fig.~\ref{fig:numerical}. We study a dense grid of initial conditions in $(X,Y)$, for $a=29,600~$km and $i =57^{\circ}$ using two numerical models: (DA) is based on the doubly-averaged formulation of Eq.~(\ref{eq:hsecular}) \cite{Gkolias2016} and (HF) is a non-averaged symplectic propagator \cite{Rosengren2019} of the complete model of Eq.~(\ref{eq:fullmodel}). A short-time Fast Lyapunov Indicator (FLI) \cite{Froeschle1997} map was computed, to depict the $(X,Y)$-projection of the stable manifolds emanating from the NHIM \cite{Lega2016}. We then extended our integrations to $\sim 465~$years and computed maps of escape time, $T_{esc}$, defined here as the time taken for an orbit to enter the shaded area of Fig.~(\ref{fig:analytical}). The $T_{esc}$-maps are practically identical for (DA) and (HF), which reflects the quality of approximation of the mean, secular flow of Eq.~(\ref{eq:fullmodel}) by Eq~(\ref{eq:hsecular}).\\

There is a remarkable similarity between the spatial distribution of $T_{esc}$ values and the manifolds structure, depicted in the FLI map. Zooming in the low-$e$ domain of interest for actual satellites, a fractal-like \cite{Bleher1988,Aguirre2001,Moura2002,Nagler2004,Nagler2005,Altmann2013} stratification of short/long $T_{esc}$ `stripes' is seen in Fig~\ref{fig:numerical} for $e\leq 0.1$, in striking correspondence with the oscillations of the stable manifold of the NHIM \cite{Dvorak1998,Contopoulos2008,Aguirre2009,deAssis2014}. The direction of the stripes in $(X,Y)$ remains very close to the one given by the analytically computed stable eigenvector. Computing the histogram of $T_{esc}$ for this region, we find that it follows a power-law, $P(t)\sim t^{-\alpha}$, with $\alpha \approx 1.5$. Extending our grid to the whole disc of initial conditions $e<0.8$, we find $\alpha = 0.8$. These statistics are indicative of anomalous transport \cite{Dvorak1998,Zaslavsky2002}. 

\section{Discussion}

\emph{Resonance width} -- Our analytical model allows accurate estimate of the extent of the regions of hypebolicity of quasi-circular GNSS orbits around Lidov-Kozai resonances; it coincides with the range of $J_F$ values that give real eigenvalues of (\ref{eq:stabmat0}). Depending on $i_P$, resonant variations of $J_F$ can maximize the extent of this domain at Galileo altitudes; conversely, at $24,000~$km the double-resonance reinstates elliptic stability of the CM. Our approximations were validated in a series of numerical experiments. \\  

\emph{Role of lunar node regression} -- The harmonics in $\mathcal{H}_{CM,1}$ would not exist if the Moon's ecliptic inclination, $i_{\leftmoon}$, was zero -- their coefficients are proportional to $\sin{i_{\leftmoon}}$. The $\mathcal{R}_{2h-\Omega_{\leftmoon}}$ resonance occurs precisely at Galileo altitudes, because of the value of the lunar nodal precession period ($18.6~$years). Our analytical model confirms previous numerical simulations, which have attributed the chaos observed to the regression of the lunar nodes \cite{Rosengren2015,Rosengren2017}. Moreover, it predicts a chaotic region of size $\Delta e \gtrsim  0.2$ around the circular orbit, from which transport to `Earth-crossing' orbits emanates. We also explain the existence of a significant fraction of long-term stable, high-$e$ orbits (those with $u_R \approx \pm \pi/2$ in Fig.~1), as found in [\onlinecite{Skoulidou2019}]. \\

\emph{Manifold design of EoL} -- Using the hyperbolicity of the GNSS region for designing EoL trajectories, possibly in synergy with a low-cost impulsive maneuver to a sizeable eccentricity (i.e.\ $0.01-0.05$), is appealing. An e-folding time of $36~$years implies $T_{esc}\sim 120~$years for $e\sim 0.05$. The re-entry trajectory of Fig.~1 has $T_{esc} = 125~$years, as do all orbits inside the same `turquoise' strip (see Fig.~\ref{fig:numerical}). The next, nearly parallel, turquoise strip towards the origin has $T_{esc} \sim 160~$years, i.e.\ an additional e-folding time, while $T_{esc}\sim 80~$years for $e\sim 0.1$ have been found by [\onlinecite{Skoulidou2019}]. Given the fractal distribution of manifolds crossings, chaotic trajectories adjacent (but not inside) these stripes may wander inside the chaotic region for hundreds of years, without evacuating the operational zone (see Fig.~\ref{fig:analytical}).\\

Safe prediction of the re-entry time is important for EoL design but, particularly for Galileo, this is apparently hindered by the intricate manifolds structure in the double-resonance domain. Nevertheless, insightful maneuvering, guided by an accurate model of the local manifold dynamics is, in principle, feasible. In our model, this would correspond to targeting one of the turquoise strips, encircled by the stable manifold, by maneuvering along the nearly perpendicular unstable eigenvector, i.e.\ such that $u_R = \pi/4$~or~$5\pi/4$ for $\Omega_{\leftmoon} = \pi$. \\

\begin{acknowledgments}
I.G. acknowledges the support of the ERC project 679086 COMPASS ``Control for Orbit Manoeuvring through Perturbations for Application to Space Systems''. J.D. acknowledges the support of the ERC project 677793 ``Stable and Chaotic Motions in the Planetary Problem''. The authors would like to thank Gabriella Pinzari, Camilla Colombo, Martin Lara and Aaron Rosengren for useful discussions.
\end{acknowledgments}

\bibliography{meobib}

\end{document}